# Is photoisomerization of stilbene in solution directly promoted by solvent collisions?

S. A. Kovalenko,* A. L. Dobryakov, N. P. Ernsting*

*Department of Chemistry, Humboldt-Universität zu Berlin, Brook-Taylor-Str. 2,
D-12489 Berlin, Germany*

**Abstract**

Gas phase thermal isomerization rates of *trans*-stilbene at $T \geq 300$ K obey the Arrhenius law. In solution the rates increase ten-fold and depend only very weakly on excitation wavelength $\lambda_{exc}$ or intramolecular temperature. Both observations are difficult to reconcile within RRKM theory. Previously discussed mechanisms, like restricted IVR, nonadiabaticity, solvent-dependent barriers, or excitation-induced cooling, may explain the increase of the rates but not their $\lambda_{exc}$ dependence in solution. The latter dependence suggests that solvent collisions can directly promote the isomerization.

## 1. Introduction

Photoinduced isomerization of *trans*-stilbene[1-3] has been experimentally studied over 40 years under various conditions: in supersonic jets,[3-7] in low and high pressure gases,[3,8-16] and in liquids.[9,13-19] However this fundamental reaction is still not fully understood. One unresolved question is why the isomerization rate in solution exceeds the rate in low pressure gases by an order of magnitude,[12-15] whereas transition-state or RRKM theories[20-24] predict similar rates.

In the late 80s Fleming and coworkers[14-16] discussed the problem and summarized four possible sources for the discrepancy. (**i**) The reaction may be nonadiabatic in an isolated molecule but becomes more adiabatic in liquids with increasing collision frequency. (**ii**) Intramolecular vibrational redistribution (IVR) may be slow (restricted) in isolated stilbene, but becomes faster (unrestricted) in solution due to solute-solvent interactions. (**iii**) A barrier $E_b$ to isomerization may lower due to clustering with growing buffer density. (**iv**) Solute-solvent collisions may directly increase the isomerization rate.[15]

Later two more explanations were proposed: (**v**) excitation-induced cooling[24,25] and (**vi**) dynamic polarization[26,27] of stilbene molecules.

Consider these scenarios in more detail. Point (**i**) was put forward by Zewail and coworkers[3-7] to treat their jet and vapor experiments under collisionless conditions. They observed a strong dependence of the isomerization rate $k(E)$ on intramolecular energy $E = E_0 + (1/\lambda_{exc} - 1/\lambda_{00})$. Here $\lambda_{exc}$ and $\lambda_{00}$ are the excitation and 0-0 transition wavelength, and $E_0$ is the thermal energy. However when RRKM rates were calculated with a reaction mode $\nu_R$=400 cm$^{-1}$, the result was 5-10 times higher than the experimental one. One way to bring the calculations into agreement with experiment was to decrease the frequency of the reaction mode. But the authors preferred to keep $\nu_R$=400 cm$^{-1}$ as calculated by Warshel[28] since no better calculation was available at the time. Instead they concluded that nonadiabaticity slows down the reaction. It was thought that with increasing gas pressure and in liquids the reaction becomes more adiabatic and the reaction rate increases.

Troe and coworkers[11-13] chose another way (**iii**). They fitted gas-phase and jet rates $k(E)$ with an adjustable barrier height $E_b$. With the reaction frequency $\nu_R$ =25 cm$^{-1}$ and $E_b$ =1155 cm$^{-1}$ good agreement to low pressure gas-phase data was reached, but the acceleration of the reaction in liquid solutions remained unexplained.[13]

Therefore other workers[29-31] explored (**ii**), restricted IVR. Nordholm[29] proposed that the RRKM rate is effectively reduced by factor $k_{IVR}/(k_{IVR}+\nu_R)$ which depends on intramolecular energy and buffer-gas pressure. Leitner et al.[30] developed this scheme, their $k_{IVR}(E)$ agreed with experiment.[3,6] At $E$=2000 cm$^{-1}$ for example, they found $k_{IVR}(E)\sim$1 ps$^{-1}$; then with $\nu_R$ =600 cm$^{-1}$ the authors were able to model the pressure dependence of the isomerization rates. The approach was applied by Weston and Barker[31] to fit extensive pressure- and buffer-gas-dependent data by Meyer et al.[12] The conclusion was that the experimental rates $k(E)$ can be reproduced to within about a factor of 2.

Now regard (**iv**), that solute-solvent collisions may directly promote the isomerization. This view is consistent with temperature-dependent measurements of the quantum yield in solutions.[32-34] It also explains the initial linear pressure dependence of the reaction rates in buffer gases.[12,15] But RRKM theory in the high pressure limit or in solution demands the *independence* of reaction rates on solvent parameters.[21,22] Therefore a collision-induced reaction channel was not seriously discussed in the literature, and the mainstream followed the restricted IVR hypothesis.[29-31]

A different explanation (**v**) was proposed by Pollak and coworkers.[24,25] They note that 0-0 excitation should cause cooling of a stilbene molecule because of a jump in heat capacity (as the excited-state vibrational frequencies are on average lower than the ground-state ones). In the gas phase the isomerisation rate should be slowed down correspondingly. In solution, on the other hand, ambient temperature is quickly established for the excited molecule and the rate is significantly increased compared to the gas phase. The question is in the magnitude of the effect. Initially a temperature drop by 100 K was reported,[24] afterwards it was diminished to 10-20 K.[25] The latter is however too small to substantially affect the isomerization rates.

Lastly consider (**vi**), the dynamic polarization model by Hamaguchi and coworkers.[26,27] It assumes that the central C-C bond of stilbene is randomly polarized by solvent fluctuations, altering between the stable double-bond and virtual single-bond configuration. The isomerization occurs mainly in the single-bond configuration and is more probable when the fluctuations are faster. The model implies a direct influence of the solvent fluctuations (collisions) on the isomerization rate, and hence it belongs to (**iv**).

The present paper aims to discuss mechanisms for stilbene isomerization. We suggest that the role of solvent collisions can be found out from a (very weak) rate dependence $k(\lambda_{exc})$ on excitation wavelength. In solution these rates are nearly independent of $\lambda_{exc}$. Although known for long time, the significance of this fact has not been recognized. Commonly one assumes that the excess intramolecular energy $(E - E_0)$ is rapidly transferred to the solvent[35,14] so that no memory of the excitation is left. But new evidence shows that this is not so. The cooling dynamics of a molecule in solution can be well resolved with current ultrafast techniques.[36,17,18] For example in n-hexane the cooling of *trans*-stilbene occurs monoexponentially with $\tau_{cool} \approx 10$ ps.[17,18] Since the dependence $k(E)$ is well established from gas-phase RRKM work, one can now model[24] the population decay in solution at different $\lambda_{exc}$ and compare the result with experiment. No agreement is achieved with experiment even if scenarios (**i**)-(**iii**) are invoked. This will be demonstrated in Sections 3, 4. But first we analyze gas-phase literature data[3,8] to show that they are consistent with RRKM predictions, and hence with complete (unrestricted) IVR.

## 2. Arrhenius behavior in the gas phase

In jets and low-pressure gases microcanonical rates $k(E)$ are usually measured.[3-8] Of importance are also thermal rates $k(T_{int})$ calculated as[1,4,11-15,21,22]

$$k(T_{int}) = \int_{E_b}^{\infty} dE\, k(E) f(E) \tag{1}$$

where $T_{int}$ is intramolecular temperature, $f(E)$ is a thermal distribution function. One of the central RRKM results is that the rates in high pressure gases or in liquids cannot exceed $k(T_{int})$.[21,22]

The rates $k(T_{int})$ for *trans*-stilbene are shown in Fig. 1. They are derived from $k(E)$ originally measured by Zewail[3] and Hochstrasser[8] at collisionless conditions, using an approximation

$$k(T_{int}) \approx k(E), \quad E = \sum_j \frac{\nu_j}{\exp(\nu_j / k_B T_{int}) - 1} \quad (2)$$

where $\nu_j$ are the $S_1$ vibrational frequencies. The approximation (2) works well at high energies $E >> E_b$.[11,15] At lower energies, $E \approx 2000$ cm$^{-1}$, it is also correct as stilbene was prepared *thermally* at 296 K.[3] The behavior in Fig. 1 satisfies the Arrhenius law

$$k(T_{int}) = A_{int} \exp(-E_b / k_B T_{int}) \quad (3)$$

with $A_{int} = 0.5$ ps$^{-1}$, $E_b = 1320$ cm$^{-1}$. These fit parameters depend on the vibrational spectrum $\nu_j$, taken here from Negri and Orlandi.[37] Other frequency sets[38] result in $A_{int} = 0.4$-$0.6$ ps$^{-1}$ and $E_b = 1250$-$1350$ cm$^{-1}$. For comparison, Troe and coworkers[13] report $A_{int} = 0.73$ ps$^{-1}$ and $E_b = 1155$ cm$^{-1}$. Note that these results agree with RRKM theory which predicts for a low-frequency reaction mode ($\nu_R << k_B T_{int}$) the Arrhenius behavior.[4,13,21,22] This in turn is consistent with the RRKM hypothesis of complete (unrestricted) IVR, at $T_{int} \geq 300$ K.

### 3. Modeling the behavior in solution

When switching to the liquid phase, we pay attention to two points. First, RRKM theory considers $k(T_{int})$ given by (1), (3) to be the maximal rate for isomerization in solution.[4,11-13,15,21,22] The second point concerns the meaning of the Arrhenius law, $k = A \exp(-E_b / k_B T)$. Namely, the question is what temperature and prefactor, *solute* or *solvent*, enter into the expression? At equilibrium $T_{solv} = T_{int}$ the uncertainty is hidden, but generally $T_{solv} \neq T_{int}$ and the problem becomes evident. The use of solvent parameters $T_{solv}$, $A_{solv}$ seems reasonable as this agrees with the weak dependence on $\lambda_{exc}$, but it would suggest a different activation mechanism.

Alternatively one may try to use the RRKM rate (3) together with one of the assumptions (**i**)-(**iii**) to model the behavior in solution. Then the excited-state population $N(t)$ can be expressed as

$$dN/dt = -k(T_{int})N - N/\tau_F \qquad (4)$$

where $\tau_F$ is the fluorescence lifetime, and the rate $k(T_{int}) = A_{int}\exp(-E_b/k_B T_{int})$ depends on intramolecular temperature $T_{int}$, which is now time-dependent due to vibrational cooling[17,18,24]

$$T_{int}(t) = T_{solv} + (T_0 - T_{solv})\exp(-t/\tau_{cool}) \qquad (5)$$

For clarity consider *trans*-stilbene in n-hexane, in which case the vibrational cooling proceeds monoexponentially with $\tau_{cool}=10$ ps.[17,18] Now assume (**i**) nonadibaticity or (**ii**) restricted IVR (which becomes unrestricted in solution) to be responsible for the rate acceleration in solution. With $\lambda_{exc} \approx \lambda_{00}=326$ nm in n-hexane the intramolecular temperature stays unchanged (neglecting excitation-induced cooling for the moment), $T_{int}(t) \approx T_{solv}=293$ K, and the population decay is monoexponential with $\tau=82$ ps. The behavior is shown in Fig. 2 by a black curve. Using this decay one can simulate $N(t)$ for $\lambda_{exc}=267$ nm which corresponds to the initial intramolecular temperature $T_0=607$ K. Eqs. (4), (5) are solved numerically with $E_b=1320$ cm$^{-1}$ to result in $N(t)$ drawn as red curve. Such population drops by 80% within 10 ps have never been observed experimentally.

Next assume (**iii**), lowering of the barrier when going from gas to liquid. The consistency with the solution rates requires $E_b=840$ cm$^{-1}$, and simulated $N(t)$ for that case is shown in Fig. 2 by green curve. Again the early population decay is too strong compared to experiment. Thus the nonadiabaticity, restricted IVR or barrier lowering cannot reproduce the experimental $k(\lambda_{exc})$ dependence in solution.

We therefore return to (**iv**), a collision-induced channel of stilbene isomerization is assumed in addition to the intramolecular one. Accordingly Eq. (3) is modified to explicitly include the solvent contribution

$$d(\ln N)/dt = -k_{int} - k_{solv} - 1/\tau_F$$
$$= -A_{int}\exp(-E_b/kT_{int}) - A_{solv}\exp(-E_b/kT_{solv}) - 1/\tau_F \quad (6)$$

Here $k_{int}$ is the gas-phase rate (3) which is responsible for the decay due to intramolecular energy flow at temperature $T_{int}(t)$. The second term

$$k_{solv} = A_{solv}\exp(-E_b/kT_{solv}) \quad (7)$$

represents the solvent-induced isomerization rate at temperature $T_{solv}$. The prefactor $A_{solv}$ is an effective collision frequency, while the Boltzmann factor indicates that only collisions with energies higher than $E_b$ promote the isomerization. According to Eq. (6) the reaction mode interacts simultaneously with two heat baths, of temperature $T_{int}$ and $T_{solv}$. During the isomerization $T_{int}(t)$ decreases exponentially according to (5) whereas $T_{solv}$ remains constant. With $\lambda_{exc} = \lambda_{00}$ one has $T_{int} = T_{solv}$, and $A_{solv}$ can be calculated from

$$(A_{int} + A_{solv})\exp(-E_b/k_B T_{solv}) = (1/\tau - 1/\tau_F) \quad (8)$$

where $\tau$ is the experimental decay time in solution. For n-hexane $\tau$ =82 ps at $T_{solv}$=293 K, $\tau_F$ =1.6 ns, $A_{int}$ =0.5 ps$^{-1}$, $E_b$ =1300 cm$^{-1}$, $k_B T_{solv}$ =204 cm$^{-1}$, one gets $A_{solv}$ =6.3 ps$^{-1}$. It follows that $A_{solv} \gg A_{int}$, and hence the population decay depends *weakly* on $T_{int}$, or on $\lambda_{exc}$. One can substitute the obtained $A_{solv}$ into Eq. (6), solve it numerically for $\lambda_{exc}$=267 nm, and compare the result to experiment. This will be done in the next section.

**4. Transient spectra and kinetics**

Transient absorption spectra[39,18,19] of *trans*-stilbene in hexane with $\lambda_{exc}$=326 nm and 267 nm are shown in Fig. 3. When $\lambda_{exc} = \lambda_{00}$ (top) stilbene molecules stay approximately at temperature $T_{solv}$=293 K. With $\lambda_{exc}$=267 nm the molecules are initially hot, $T_0 \approx 600$ K. This is seen by broadening and red shift of the prominent band for excited-state absorption (ESA), compared to late time. As the molecules are cooled down by the solvent, the band becomes narrower and shifts to the blue. To cancel these effects, the signal decays are measured by the band integral over 408-690 nm.[18]

The difference between the two population decays (with $\lambda_{exc}$=326 nm and 267 nm) is small. Therefore cautions were taken to reduce systematic errors and to improve the signal-to-noise. The room temperature was kept constant to ensure $T_{solv}$=(20.3±0.3) C°. Pump-probe scans at the magic angle with 6, 40, 200 and 1000 fs steps were recorded from negative-to-positive and from positive-to-negative delays, with averaging over 20-40 scans.

Fig. 4 shows the decay kinetics. The difference signal (magnified by 5, red symbols) is fitted to Eq. (6). The best fit gives $E_b$=1361 cm$^{-1}$, $A_{solv}$=8.7 ps$^{-1}$ for hexane; and $E_b$=1248 cm$^{-1}$, $A_{solv}$=10.6 ps$^{-1}$ for acetonitrile. Small deviations at early time (~10 ps) may be due to solvent heating around the solute.[18] While this heating is weak, it may contribute because of the factor $\exp(-E_b/k_B T_{solv})$. Note that the fitted barrier height in hexane agrees with that obtained from standard temperature-dependent measurements in this solvent.[1,14,32-34] Also, $A_{solv}$~10 ps$^{-1}$ is close to theoretical estimates for the collision frequency.[40,10]

## 5. Discussion

In the gas phase at collisionless conditions, our results are close to those of Troe and coworkers.[13] They are consistent with the RRKM theory[4,13,20-24] which predicts for a low-frequency reaction mode ($\nu_R \ll k_B T$) the Arrhenius behavior

$$k_{int} = \nu_R \exp(-E_b/k_B T_{int}) \qquad (9)$$

Here $\nu_R$=17 cm$^{-1}$ corresponds to $A_{int}$=0.5 ps$^{-1}$ in Eq. (3). It has already been mentioned that first RRKM rates $k(E)$ were calculated[3] with $\nu_R$=400 cm$^{-1}$. Those $k(E)$ were (by chance) similar to the rates in solution, providing support to the view that $k(T_{int})$ should be the upper limiting rates, and simultaneously indicating that something might be different with the gas-phase rates. This gave rise to the assumptions of nonadibaticity,[3] restricted IVR,[29-31] and excitation-induced cooling.[24,25]

In buffer gases and in liquid solution, we propose that solute-solvent collisions provide a new activation channel in addition to the intramolecular activation. The full isomerization rate is given by Eq. (6), thus capturing both the rate acceleration in solution and the weak dependence on $\lambda_{exc}$. Previous attempts, such as restricted IVR,[29-31] nonadiabaticity,[3] pressure-dependent barriers,[13] or excitation-induced cooling[24,25] may explain only the increase of solution rates but fail to predict the correct $\lambda_{exc}$ dependence.

The view that solute-solvent collisions may affect the reaction rate is not new. Let us discuss the well-known Lindemann scheme[22,41,42] which predates RRKM theory and is contained in it. In this scheme the reaction is activated by collisions and described as

$$A + M \xrightarrow{k1} A^E + M \quad \text{(activation)}$$
$$A^E + M \xrightarrow{k2} A + M \quad \text{(deactivation)} \quad (10)$$
$$A^E \xrightarrow{k(E)} B \quad \text{(isomerization)}$$

where $M$ is a buffer gas ($[M] \gg [A], [A^E]$), and $A$, $A^E$ and $B$ are the educt, excited educt and product; $k_1$, $k_2$ are activation and deactivation rates, and $k(E)$ is the isomerization rate for molecules with energy $E$. Note that $k(E)$ is the RRKM rate given by Eqs. (1)-(3) or (9). Under steady-state condition for $[A^E]$ the unimolecular rate becomes[22]

$$k_{uni} = \frac{k(E) \cdot k_1 [M]}{k(E) + k_2 [M]} \quad (11)$$

At low pressure $k_{uni}$ is proportional to $[M]$ or to pressure, while at high pressures and in solution one obtains

$$k_{uni} = \frac{k(E) k_1}{k_2} \approx k(E) \quad (12)$$

This is the well-known RRKM result that $k_{uni}$ cannot be higher than the RRKM rate $k(E)$.[21,22] However experimental high-pressure rates exceed[15] $k(E)$ for any reasonable choice of $\nu_R$, up to $\nu_R = 400$ cm$^{-1}$. To circumvent this problem Troe and coworkers introduced the pressure-dependent barriers,[12] while Nordholm[29] and others[30,31] explored the idea of restricted IVR. But according to Section 3, these mechanisms as well as nonadiaticity, or excitation-induced cooling disagree with the experimental $\lambda_{exc}$ dependence in solution.

One way to resolve the problem is to assume that collisions not simply energize a molecule but directly excite the reactive mode to bring the molecule to the top of the reaction barrier or higher. This is the view which is tentatively proposed here. Then $k(E)$ in (10), (11) should be substituted by $k_3$, the *barrierless* isomerization rate. As $k_3 \gg k(E)$, the linear dependence $k_{uni} = k_1[M]$ is preserved from low through high pressure gases up to the liquid phase. Recent experiments[43] on α-substituted stilbenes indicate $k_3 \sim 10$ ps$^{-1}$, providing support to this scenario. We are aware that such a mechanism violates the RRKM assumption on the intramolecular energy redistribution.

A realization of the above mechanism could be collisions with momentum projection perpendicular to the molecular plane of stilbene. Such collisions excite the phenyl motion just along the reaction coordinate. The probability for solvent molecules to have energy $E > E_b$ is $\exp(-E_b/kT_{solv})$, and one recovers the rate (7), at the condition that *all* the energy $E$ is transferred to the reactive mode. If however only a part of the energy, $\alpha \cdot E$ ($\alpha < 1$), were transferred, the apparent barrier would *increase* corresponding to $E_b/\alpha$. In a very crude approximation one can use the results from elastic collisions of particles. In that case the complete energy transfer occurs for equal masses. Hence one may expect that the collisional energy transfer will be efficient for hexane or pentane solvents as their molecular masses are similar to that of the phenyl unit. On the contrary, for light solvent molecules like hydrogen or helium the energy transfer should be inefficient. Qualitatively this picture agrees with the data of Ref. 12, which show permanent increase of the isomerization rates in the series He, Ne, Ar, Xe, and in the series $CH_4$, $CO_2$, $C_2H_6$, $C_3H_8$. However, detailed molecular dynamic simulations are necessary to show thoroughly if the present activation mechanism indeed agrees with experiment.

## 6. Conclusion

Various isomerization mechanisms of photoexcited stilbene were discussed. It was shown that nonadiabaticity, restricted IVR, solvent-dependent barriers, or excitation-induced cooling cannot be reconciled with the experimental $\lambda_{exc}$ dependence of the reaction rates in solution. Instead an additional reaction channel was proposed to bring the rates into agreement with experiment. This collisional activation may be realized either with direct excitation of the phenyl motion along the isomerization coordinate, or by dynamic polarization of the central C-C bond. Further studies are necessary to clarify which activation mechanism is operative.

**Acknowledgement**

We thank Professors M. Maroncelli, J. Troe, J. Schroeder, D. Schwarzer and E. Pollak for many helpful comments, and the Deutsche Forschungsgemeinschaft for financial support.

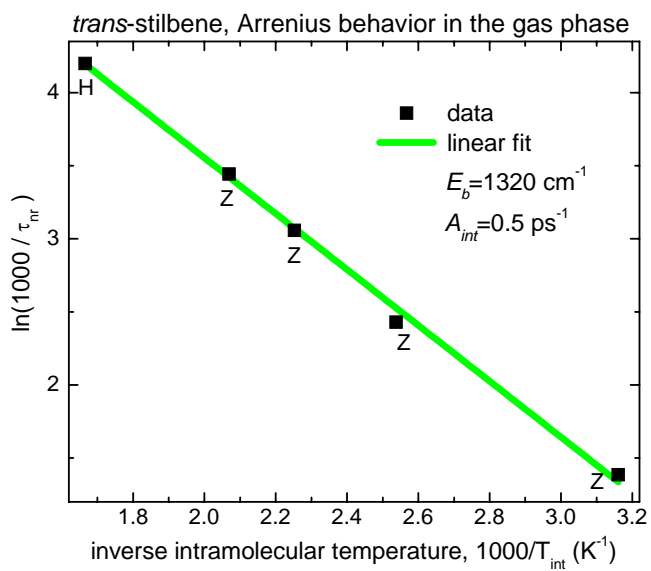

**Fig. 1.** Gas-phase thermal isomerization rates $k(T_{\text{int}}) = 1/\tau_{nr}$ ($\tau_{nr}$ is nonradiative decay time in ps) at collisionless conditions satisfy the Arrhenius law (3) with $A_{\text{int}} = 0.5$ ps$^{-1}$, $E_b = 1320$ cm$^{-1}$. The data are from Zewail[3] (Z) and Hochstrasser[8] (H). *trans*-Stilbene molecules were prepared at 296 K (Z), or at 390 K (H).

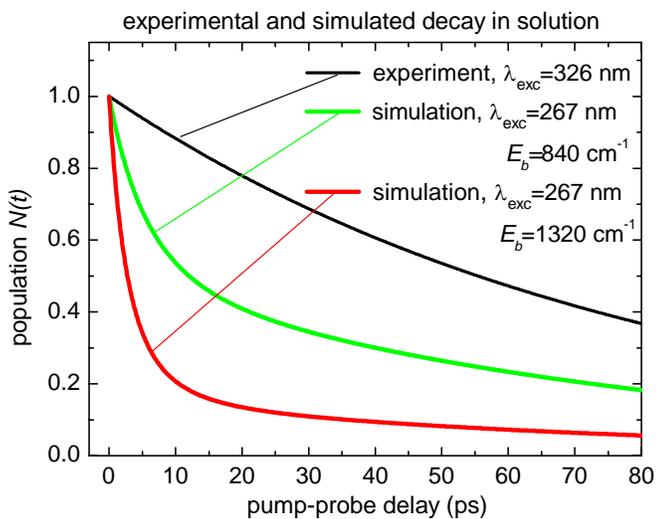

**Fig. 2.** Black curve is experimental excited-state population decay (monoexponential, $\tau = 82$ ps) of *trans*-stilbene in hexane at $T_{solv}=293$ K and $\lambda_{exc}=326$ nm (zero excess energy). Red and green curves show simulated decays from Eqs. (4), (5) with $\lambda_{exc}=267$ nm. The red curve corresponds to mechanisms **(i)** or **(ii),** if they were responsible for the rate acceleration in solution. The green curve corresponds to **(iii).** The simulated behaviors have never been experimentally observed (see Fig. 4).

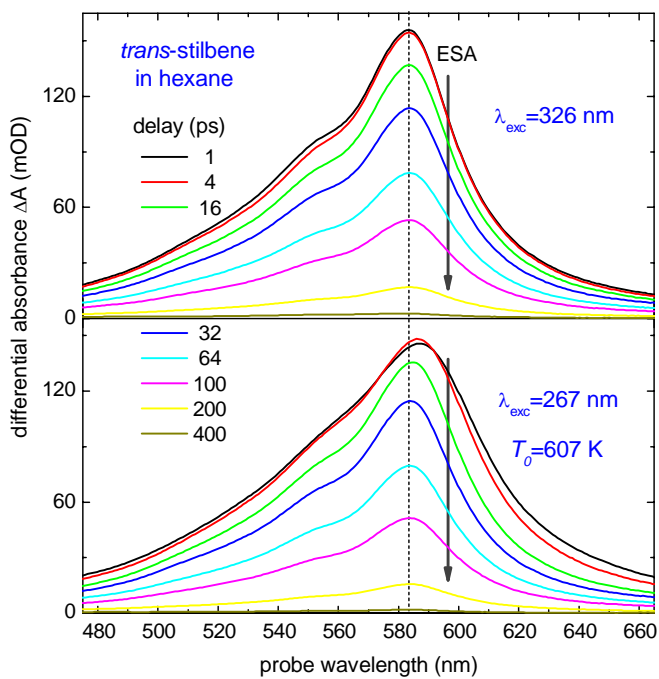

**Fig. 3**. Transient absorption spectra[18,19] of *trans*-stilbene in hexane at $T_{solv}$ =293 K, with $\lambda_{exc}$ =326 nm (top) and 267 nm (bottom). In the first case $\lambda_{exc} = \lambda_{00}$, and intramolecular temperature stays approximately constant, $T_{int} = T_{solv}$. With $\lambda_{exc}$ =267 nm the molecule is initially hot, $T_0$ =607 K. When it cools down the ESA band becomes narrower and shifts to the blue. These effects are eliminated by integrating the decay kinetics over 408-690 nm.[18]

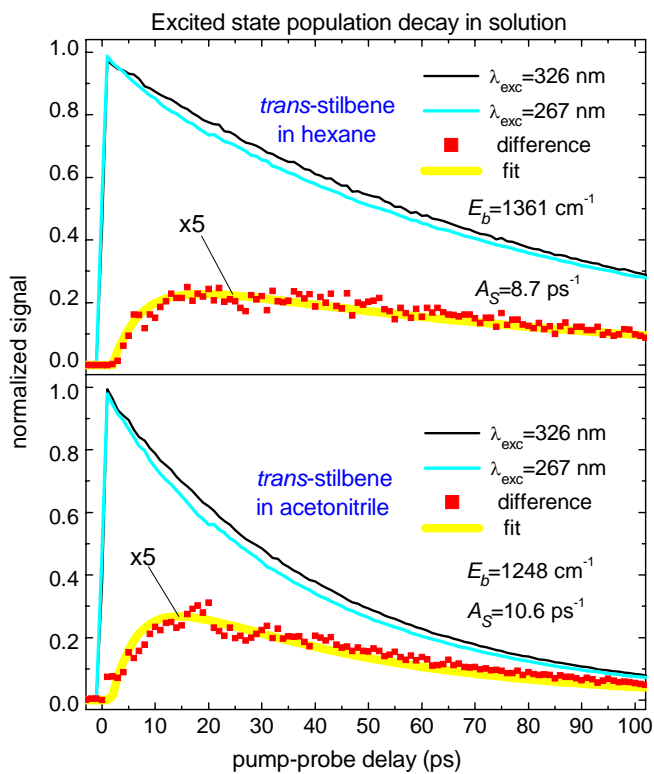

**Fig. 4.** Excited-state decay of *trans*-stilbene in solution at $T_{solv}$ =293.3 K. With zero excess energy ($\lambda_{exc}$ =326 nm, black curve) the decay is monoexponential, $\tau$ =82 ps in hexane, and $\tau$ =40 ps in acetonitrile. With $\lambda_{exc}$ =267 nm (cyan) the decay is similar but a bit faster. The difference signal (magnified by 5, red symbols) is fitted to Eq. (6) (yellow curve) giving $E_b$ =1361 cm$^{-1}$, $A_{solv}$ =8.7 ps$^{-1}$ for hexane, and $E_b$ =1248 cm$^{-1}$, $A_{solv}$ =10.6 ps$^{-1}$ for acetonitrile.